# A Visual Model for Web Applications Security Monitoring

Tran Tri Dang, Tran Khanh Dang
*Ho Chi Minh City University of Technology*
tridang@cse.hcmut.edu.vn , khanh@hcmut.edu.vn

**Abstract**

*This paper proposes a novel visual model for web applications security monitoring. Although an automated intrusion detection system can shield a web application from common attacks, it usually cannot detect more complicated break-ins. So, a human-assisted monitoring system is an indispensable complement, following the "Defense in depth" strategy. To support human operators working more effectively and efficiently, information visualization techniques are utilized in this model. A prototype implementation of this model is created and is used to test against a popular open source web application. Testing results prove the model's usefulness, at least in understanding the web application structure.*

## 1. Introduction

With the increasing need for new e-commerce sites and online services, there is also an increasing need in creating more web applications for that purpose. These web applications are created not only by professional companies but also by individuals with more or less experience. It is clear that not everyone involving in web development understands about the nature of security issues in their works and that make those applications result in a poor security setting. Although web applications are widely deployed and usually accessible from the outside world, to the best of our knowledge, there is no security monitoring tool created specifically for web applications.

In this paper, we propose a visual model for monitoring security related issues of web applications. The model uses existing structures of web applications as the base for its monitoring mechanism. This is based on the fact that when developing a web application, the developers need to decide in advance the structure of their application to which they expect the visitors will obey when using the application. If there is an access violating this structure, it is abnormal and further investigations are needed to verify whether it is a step in a real attack. In addition, the proposed model is visual so it can take advantage of visual recognition capability of human vision and perception. Visual display can also present more information at a time than text display. Finally, user interaction is often more intuitive in the visual environment.

The rest of this paper is structured as follows: in Section 2, we present related works; in Section 3, our visual model is described; in Section 4, the prototype implementation is described; Section 5 presents some experiments and results; and Section 6 concludes the paper with future works.

## 2. Related works

There are different approaches in the field of web application security. Scott and Sharp [12] propose a system to harden web applications. That system positions between a web server and users, examining users' inputs with regards to the security specifications written in the system's own language. It is capable of handling popular attack types on web applications like form modification, SQL injection and Cross-Site Scripting (XSS). However, operators have to write security specification code manually, which will be a complex task for large applications. Huang et al. [4] propose a system capable of detecting two common types of security vulnerabilities in existing web applications: SQL injection and XSS. For SQL injection vulnerability detection, the system crafts malicious inputs and sends to the server for processing; it then observes the response text, looking for suspected strings, e.g. "ODBC error", that the application may produce. For XSS vulnerability detection, the web application is initial crawled in learning mode to study the normal behavior of the crawler. That behavior is then used to compare to crawler's behaviors in detection mode. If they do not match, the related pages are considered to contain injected code and hence have XSS vulnerabilities. For this reason, the system can only detect XSS vulnerabilities on a web application once that application is attacked successfully. A similar work to [4] is introduced by SecuBat [5]. However, SecuBat can detect reflected XSS vulnerabilities early by inserting simple JavaScript code and checking response text.

Besides prevention approaches described above, Kruegel and Vigna [6] propose a detection approach in which each query to a web application is examined in various models for calculating its anomaly score. Each model handles a different input aspect, e.g. attribute length, attribute structure etc. There are two modes in which each model can operate: training mode and detection mode. The training mode is used to learn the characteristics of normal events and to calculate anomaly score thresholds between regular and anomalous inputs. After the training mode is finished, the detection mode is used to calculate anomaly scores and report anomalous queries. Although the work result of [6] can be used to detect certain kinds of web application attacks, its text-based reports are difficult to comprehend and require further effort of operators to do in-depth investigation for suspected cases.

In the field of web visualization, there are researches aiming at effectively visualizing web space for the purposes of understanding its structure, supporting discovery process, enhancing usability and providing relevant search results. Munzner and Burchard [8] use a 3D Hyperbolic space to visualize the structure of a web space to help web surfers have an insight about the overall relationship between pages. To display more information on a restricted screen, Hyperbolic space is used because of its superior ability in this case compare to Euclidian space's. Dachselt and Ebert [2] propose a Collapsible Cylindrical Trees technique for fast and intuitive interaction on hierarchical structures, e.g. web site navigation. The navigation is very fast with just only one click is needed to select an action. Weinreich and Lamersdorf [15] use hyperlinks visualization to help web surfers understand more about a link before clicking on it, making the web surfing experience more comfortable. Ortega and Aguillo [9] use link analysis and visualization to learn about Nordic academic web network. The resulting work helps recognizing sub-networks in the main network, understanding the role of each university web space and the relationships between them. Chung et al. [1] apply visualization techniques to the results returned from meta-search engines to support the knowledge discovery process. Its visual presentation provides three different types of result browsing mode for different purposes: sequential display, grouping and tree display, and scatter plot display.

Although security visualization is an active research field with a divergent application area, e.g. from network intrusion detection [7] to peer-to-peer resource sharing applications [3], as far as we know, there is currently no published work on visualization for web application security.

## 3. A visual model for web applications security monitoring

In web application security, managing user inputs is a crucial part. Among the popular attack types on web applications, the most typical ones are: form modification, SQL injection, XSS, Cross Site Request Forgery, and Insecure Direct Object References. Notably, all of them exploit input handling flaws in the application code. Because web applications are usually accessible to the public, anyone (either human or computer program) can access these applications and input anything into them for processing. HTML forms [14] are standard containers for these input values and are created by web application developers. Developers often put restrictions (by using HTML or JavaScript) on those forms to limit the possible ranges and formats of these inputs' values. In this case, developers make an assumption that the inputs will be constrained by these restrictions, which may be wrong for inputs sent from a dedicated malicious attacker, who knows how to manipulate the forms' structures to remove these restrictions. Therefore, a good web application security monitoring system should enable operators effectively and efficiently recognizing the deviations to the existing forms' structures in users' requests.

Our visual model for security monitoring displays events happening on forms with a concentration on forms' structures. To overcome the information overload problem, this model supports different levels of details and the "visual information seeking mantra" interaction proposed by Shneiderman [13], i.e. "overview first, then zoom and filter, and finally, details on demand". In our model, there are three levels of abstraction, each has a different visual design and provides particular information in that level. To get more detailed information, operators can go down one level, and to get more summarized information, operators can go up one level. In addition, to help operators in the discovery process, we use a different color to display in advance the abnormal events existed in lower levels so that operators can go down to look for more information. This technique is borrowed from [11], which is originally used in navigating on an OLAP data cube.

The model's levels, from the highest to lowest abstraction, are briefly described below:

- At the highest abstraction level, a collection of forms with the same destination page is treated at a whole. These forms are distributed equally around a circle whose center is the destination page. Then, each normal form request with specified destination page will be mapped to the appropriate form area according to its values. One additional dummy form area is also created to contain abnormal requests at this level (for requests that are normal at this level, but abnormal at a lower level, we still map them at the right location but with a different color). That area is recognized easily by its different color. To be considered normal at this level, form requests must contain valid methods and are sent from particular URLs, which we call source pages. In HTML language, source pages are documents that contain form declarations and destination pages are pages whose URLs are contained in "action" attribute of the forms. Note that there is a many-to-many relationship between source pages and destination pages. But because the destination pages are the places where user inputs are actually handled, our model is destination page–oriented, i.e., we group forms by a common destination page as depicted in Figure 1.

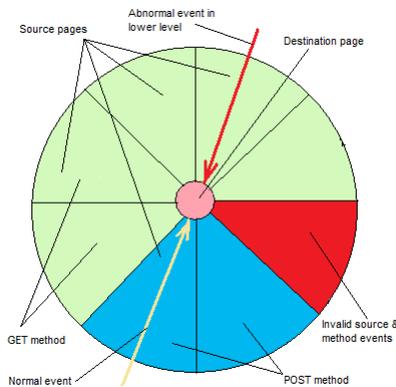

**Figure 1. Collection of forms level**

-At the second abstraction level, each form is treated individually and they can be reached from the top level by clicking on related form request. In this level, we model forms as containers containing an ordered set of controls. This set is created by application developers with an assumption of what controls are mandatory, what controls are optional and the order imposed on them. To represent this order set, we use a vertical directed line with horizontal line segments on it. Each line segment is used to present one control, with continuous segments are for mandatory controls and broken segments are for optional controls. The segments' positions are set according to controls' orders on forms. Each form request is also mapped as a vertical directed line, with each {name, value} pair contained in the request is presented as a line segment. The form line and request line are drawn parallel to each other to enable easy comparing. By examining it, we can recognize whether some mandatory controls are removed or whether some invalid controls are added. Again, a normal request {name, value} pair but is abnormal at a lower level is also presented with a different color. The visual design for this level with one normal and one violation case is presented in Figure 2.

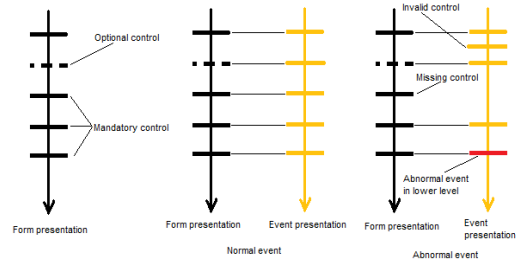

**Figure 2. Individual form level**

-At the lowest abstraction level, each form control is treated individually. Each control has its own restrictions on the possible values it can contain. These restrictions may be imposed implicitly by controls' types (e.g. a hidden control is almost read-only) or explicitly by developers (e.g. setting the maximum length attribute of a text field to restrict its value). Controls' properties, including names, types, and values, are presented in rectangles. Controls' restrictions (or constraints) are presented in ellipses around the respective rectangle. Contraints that are satisfied are presented in green, and constraints that are violated are presented in red. The visual design for this level is depicted as shown in Figure 3.

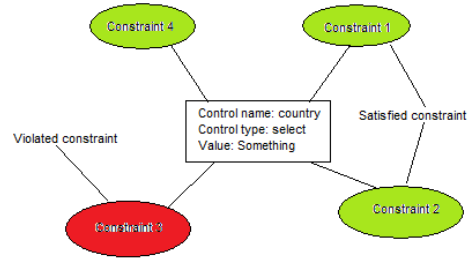

**Figure 3. Individual control level**

## 4. Prototype implementation

Our prototype implementation has two components:
-The first component is a crawler whose main task is to scan through pages, extract forms, form controls and restrictions on them. According to our model described above, it needs to know about forms' sources, methods, control sets and individual controls to enable monitoring effectively. This step is used to automatically extract these structures of any web applications. In our current prototype implementation, only HTML restrictions are recorded, JavaScript restrictions are scheduled to be recorded in later implementations due to its complexity. Another improvement will also be made in the future is to add the crawler's ability to attach labels to controls, in addition to controls' names, in order to help operators understand what values are expected to be contained in each control. Because names are intended to be processed on server, their values may not be descriptive enough, as opposed to labels which are used to guide human in input process. The problem of attaching labels to controls is mentioned in [10].

-The second component is the visualization system. Its major role is to extract security structures recorded previously by the crawler and combine them with the access events generated by web server to render visual models on computer screens. The visual models are described previously in Section 3. This component is also responsible for handling user interactions.

## 5. Experiments and results

To test our prototype implementation, we choose WordPress [16], a popular open source blogging application, and install it locally. Parts of the security structures of WordPress are visualized in Figure 4 and Figure 5.

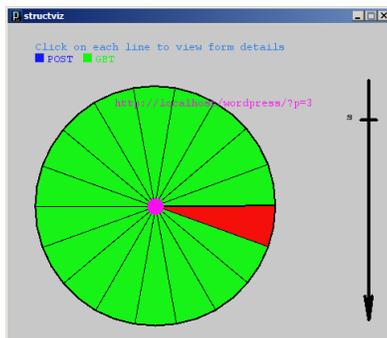

**Figure 4. Security structure for search form**

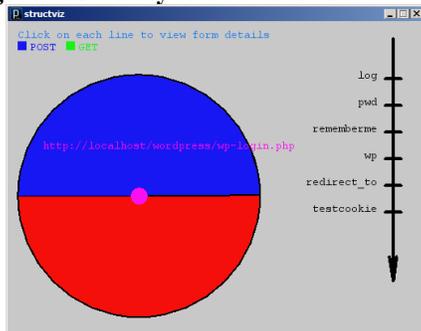

**Figure 5. Security structure for login form**

From the prototype results, we can draw some conclusions about WordPress: (1) different forms pointing to the same target have the same structure (Figure 4); (2) only search form uses GET method, other forms use POST method; and (3) some control names in WordPress are not very descriptive (e.g. 'wp' is used for submit button in login form). Also, from the results, we can see that before logged in, an attacker has at least four "doors" to attack a WordPress application (i.e. search, comment, login and lost password forms). Not all four doors have the same complexity to break in, e.g. there is only one input to handle in the search form, but there are as much as six inputs to handle in the login form (Figure 5). The implication is that the more inputs to handle, the more vulnerable a form may be, because it needs just one incorrectly handled input to break the processing code.

We plan to do more intensive experiments to verify the model's effectiveness in recognizing attacks. At the moment, this tool can only be used to understand and discover web applications security structures, as well as to demonstrate our newly proposed visual model.

## 6. Conclusions and future works

In this paper, we have proposed a visual model that can be implemented to monitor web applications for security events. In particular, it can help recognizing modifications to existing application structures, which maybe the first steps in some actual attacks. When implementing the prototype, we also found some problems that need further investigation:

-Currently, we only take into account static structure attributes, i.e. structures imposed by HTML code. There are also dynamic structures created by JavaScript and Ajax that are more complex to handle.

-The structure of a web application is an evolving entity. It has been constantly changed by developers. The crawling algorithm needs to be aware of that and capture only new changes for optimized performance.

-As the first step, the introduced visual model only proves useful in recognizing form modification attacks, so its capability is somewhat limited. It should be improved to help recognizing abnormal contents in form controls and to help connecting related events to learn more about complicated attacks.

## 7. References

[1] W. Chung, H. Chen, and J. Nunamaker Jr., "A Visual Framework For Knowledge Discovery on The Web: An Empirical Study of Business Intelligence Exploration", *JMIS*, 21(4), 2005, 57-84.

[2] R. Dachselt, and J. Ebert, "Collapsible Cylindrical Trees: A Fast Hierarchical Navigation Technique", *Proceedings of the IEEE Symposium on Information Visualization 2001*, IEEE, CA, USA, 2001, pp.79-86.

[3] T.T. Dang, and T.K. Dang, "Security Visualization for P2P Resource Sharing Applications", *IJCSE*, 1(2), 2009, 47-55.

[4] Y-W. Huang, S-K. Huang, T-P. Lin, and C-H. Tsai, "Web Application Security Assessment by Fault Injection and Behavior Monitoring", *Proceedings of the 12th International*

Conference on World Wide Web, ACM, NY, USA, 2003, pp. 148-159.

[5] S. Kals, E. Kirda, C. Kruegel, and N. Jovanovic, "SecuBat: A Web Vulnerability Scanner", *Proceedings of the 15th International Conference on World Wide Web*, ACM, NY, USA, 2006, pp. 247-256.

[6] C. Kruegel, and G. Vigna, "Anomaly Detection of Web-based Attacks", *Proceedings of the 10th ACM Conference on Computer and Communications Security*, ACM, NY, USA, 2003, pp. 251-261.

[7] Y. Livnat, J. Agutter, S. Moon, R. F. Erbacher, and S. Foresti, "A Visualization Paradigm for Network Intrusion Detection", *Proceedings of the 6th Annual IEEE SMC Information Assurance Workshop*, IEEE, NY, USA, 2005, pp. 92-99.

[8] T. Munzner, and P. Burchard, "Visualizing the Structure of The World Wide Web in 3D Hyperbolic Space", *Proceedings of the 1st Symposium on Virtual Reality Modeling Language*, ACM, NY, USA, 1995, pp. 33-38.

[9] J. L. Ortega, and I. F. Aguillo, "Visualization of The Nordic Academic Web: Link Analysis Using Social Network Tools", *Information Processing & Management*, 44(4), 2008, 1624-1633

[10] S. Raghavan, and H. G. Molina, "Crawling the Hidden Web", *Proceedings of the 27th International Conference on Very Large Data Bases*, Morgan Kaufmann, CA, USA, 2001, pp. 129-138

[11] S. Sarawagi, R. Agrawal, and N. Megiddo, "Discovery-Driven Exploration of OLAP Data Cubes", *Springer LNCS*, 1377, 1998, 168-182

[12] D. Scott, and R. Sharp, "Abstracting Application-Level Web Security", *Proceedings of the 11th International Conference on World Wide Web*, ACM, NY, USA, 2002, pp. 396-407

[13] B. Shneiderman, "The Eyes Have It: A Task by Data Type Taxonomy for Information Visualizations", *Proceedings of the IEEE Symposium on Visual Languages*, IEEE, CO, USA, 1996, pp. 336-343

[14] W3C, http://www.w3.org/TR/html4/interact/forms.html, accessed April, 2011

[15] H. Weinreich, and W. Lamersdorf, "Concepts for Improved Visualization of Web Link Attributes", *Computer Networks*, 33(1-6), 2000, 403-416

[16] WordPress, http://wordpress.org, accessed April, 2011